\documentclass{emulateapj}

\usepackage{psfig}
\usepackage{apjfonts}
\begin{document}

\def\psrone{PSR~J1928+1746}
\def\psrtwo{PSR~J0628+09}
\def\psrthree{PSR~J1906+07}
\newcommand{\be}{\begin{eqnarray}}
\newcommand{\ee}{\end{eqnarray}}
\newcommand{\Smin}{S_{\rm min}}
\newcommand{\Dmax}{D_{\rm max}}

\shorttitle{Arecibo ALFA Pulsar Survey}
\shortauthors{Cordes et al.}

\title{Arecibo Pulsar Survey Using ALFA. I. Survey Strategy
	and First Discoveries } 

\author{
  J.~M.~Cordes,\altaffilmark{1}
  P.~C.~C.~Freire,\altaffilmark{2}
  D.~R.~Lorimer,\altaffilmark{3}
  F.~Camilo,\altaffilmark{4}
  D.~J.~Champion,\altaffilmark{3}
  D.~J.~Nice,\altaffilmark{5}
  R.~Ramachandran,\altaffilmark{6}
  J.~W.~T.~Hessels,\altaffilmark{7}
  W.~Vlemmings,\altaffilmark{1}
  J. van Leeuwen,\altaffilmark{8}
  S.~M.~Ransom,\altaffilmark{9}
  N.~D.~R.~Bhat,\altaffilmark{10}
  Z.~Arzoumanian,\altaffilmark{11}
  M.~A.~McLaughlin,\altaffilmark{3}
  V.~M.~Kaspi,\altaffilmark{7}
  L.~Kasian, \altaffilmark{8}
  J.~S.~Deneva,\altaffilmark{1}
  B.~Reid,\altaffilmark{5}
  S.~Chatterjee,\altaffilmark{12}
  J.~ L.~Han,\altaffilmark{13}
  D.~C.~Backer,\altaffilmark{6}
  I.~H.~Stairs,\altaffilmark{8}
  A.~A.~Deshpande\altaffilmark{2}
  and
  C.-A.~Faucher-Gigu\`ere\altaffilmark{7}
}
\altaffiltext{1}{Astronomy Department and NAIC, Cornell University, Ithaca, NY 14853}
\altaffiltext{2}{National Astronomy and Ionosphere Center, Arecibo Observatory, HC3 Box 53995, PR 00612}
\altaffiltext{3}{University of Manchester, Jodrell Bank Observatory, Macclesfield, Cheshire, SK11~9DL, UK}
\altaffiltext{4}{Columbia Astrophysics Laboratory, Columbia University, 550 West 120th Street, New York, NY~10027}
\altaffiltext{5}{Department of Physics, Princeton University, P.O. Box 708, Princeton, NJ 08544}
\altaffiltext{6}{Department of Astronomy, University of California, Berkeley, CA 94720-3411}
\altaffiltext{7}{McGill University Physics Department, Montreal, QC H3A 2T8, Canada}
\altaffiltext{8}{Department of Physics \& Astronomy, University of
British Columbia, 6224 Agricultural Road, Vancouver, B.C. V6T 1Z1, Canada}
\altaffiltext{9}{National Radio Astronomy Observatory, Edgemont Rd, Charlottesville, VA 22903}
\altaffiltext{10}{Massachusetts Institute of Technology, Haystack Observatory, Westford, MA~01886}
\altaffiltext{11}{Universities Space Research Association/EUD, Code 662, NASA Goddard Space Flight Center, Greenbelt, MD 20771}
\altaffiltext{12}{Harvard/Smithsonian Center for Astrophysics, 60 Garden St.,
Cambridge, MA~02138}
\altaffiltext{13}{National Astronomical Observatories, Chinese Academy of
Sciences, A20 DaTun Road, Chaoyang District, Beijing, 100012, China}

\begin{abstract}
We report results from the initial stage of a long-term pulsar survey
of the Galactic plane using the Arecibo L-band Feed Array 
(ALFA), a seven-beam receiver operating at 1.4~GHz with 0.3~GHz bandwidth, 
and fast-dump digital spectrometers. The search targets low Galactic
latitudes, $\vert b \vert \lesssim 5^{\circ}$, in the accessible
longitude ranges, $32^{\circ}\lesssim \ell \lesssim 77^{\circ}$
and $168^{\circ}\lesssim \ell \lesssim 214^{\circ}$. 
The instrumentation, data processing, initial survey
observations, sensitivity, and database management are described. 
Data discussed here were collected over a 100 MHz passband centered
on 1.42 GHz using a spectrometer that recorded 256 channels every
64 $\mu$s. 
Analysis of the data with their full time and frequency resolutions
is ongoing.  Here, we report the results of a 
preliminary, low-resolution analysis 
for which the data were decimated to speed up the processing.
We have detected 29 previously known pulsars and discovered
11 new ones.  One of these, \psrone, with a period of 69~ms and a
relatively low characteristic age of 82~kyr, is a plausible candidate
for association with the unidentified EGRET source 3EG~J1928+1733. 
Another, \psrthree, is a non-recycled pulsar in a relativistic
binary with orbital period of 3.98 hr.
In parallel with the periodicity analysis, we also search the data for
isolated dispersed pulses. This technique has resulted in the
discovery of \psrtwo, an extremely sporadic radio
emitter with a spin period of 1.2~s. 
Simulations we have carried out indicate that 
$\sim 1000$ new pulsars will be found in our ALFA survey. 
In addition to providing a large sample for use 
in population analyses and for
probing the magnetoionic interstellar medium, the survey maximizes the
chances of finding rapidly spinning millisecond pulsars and pulsars in
compact binary systems. Our search algorithms will exploit the
multiple data streams from ALFA to discriminate between radio
frequency interference and celestial signals, including pulsars and
possibly new classes of transient radio sources.
\end{abstract}


\section{Introduction}\label{sec:intro} 

\setcounter{footnote}{0}
Radio pulsars continue to provide unique opportunities for testing
theories of gravity and probing states of matter otherwise
inaccessible \citep{sta03, k+04}.  In large samples, they also allow
detailed modeling of the magnetoionic components of the interstellar
medium \citep[e.g.,][]{cl02, han04} 
and the Galactic neutron
star population \citep{lbdh93,acc02}.

For these reasons, we have initiated a large-scale pulsar survey that
aims to discover rare objects especially suitable for their physical
and astrophysical payoffs.  Of particular importance are pulsars in
short-period relativistic orbits, which serve as important tools for
testing gravitational theories in the strong-field regime. Our survey
parameters and data processing are also designed to find 
millisecond pulsars (MSPs). 
MSPs with ultrastable spin rates 
can be used as detectors of long-period ($\gtrsim$
years) gravitational waves \citep[e.g.,][]{lb01,wl03,j+04}, 
while sub-millisecond pulsars (if they exist)
probe the equation of state of matter at densities
significantly higher than in atomic nuclei. 
Long-period pulsars ($P \gtrsim 5$~s) and
pulsars with high magnetic fields are also of interest with regard to
understanding their connection, if any, with magnetars~\citep{wt05},
and improving our understanding of the elusive pulsar radio emission
mechanism. Additionally, pulsars with especially large space
velocities, as revealed through subsequent astrometry, will help
constrain aspects of the formation of neutron stars in core-collapse
supernovae \citep[e.g.,][]{l+01}.  
Finally, multiwavelength analyses of particular objects
will provide further information on how neutron stars interact with
the interstellar medium, on supernovae-pulsar statistics, and on
the relationship between high-energy and radio emission from 
neutron stars.

The new survey is enabled by several innovations. First is the Arecibo
L-band Feed Array\footnote{\url{http://alfa.naic.edu.}}  (ALFA), a
seven-beam feed and receiver system designed for large-scale surveys
in the 1.2--1.5 GHz band. 
The 1.4~GHz operating frequency of ALFA is particularly well suited
for pulsar searching of the Galactic plane.  Lower frequencies suffer
the deleterious effects of pulse broadening from interstellar
scattering, while pulsar flux densities typically are much reduced at
higher frequencies.  
ALFA was constructed at the Australia
Telescope National Facility (ATNF) and installed in 2004 April at the
Gregorian focus of the Arecibo telescope.  The ALFA frontend is
similar to the 13-beam system used on the Parkes telescope for surveys
of pulsars and H{\sc i}.  The Parkes multibeam (PMB) pulsar survey of
the Galactic plane \citep{m+01,m+02,k+03,h+04,fsk+04} has been
extremely prolific, yielding over 700 new pulsars in the past seven
years. Our survey will complement the PMB survey in its sky coverage
and will exploit the much greater sensitivity of the Arecibo
telescope. However, because of the smaller size of the ALFA beams
compared to the PMB system, many more pointings
must be done to cover the same area of sky. In addition to providing
better near-term localization of pulsars on the sky, the sensitivity
of the telescope greatly decreases the time spent per pointing, which
results in much better sensitivity to pulsars in compact binary systems
without searching a large grid of acceleration values to combat
binary motion.

Second, our initial and next-generation spectrometer systems have much
finer resolution in both time and frequency than the spectrometer used
with the PMB, increasing the detection volume of MSPs by an order of
magnitude. This comes at the cost of an increase in data rate by two
orders of magnitude ($\sim 0.3$~TB~hr$^{-1}$), requiring substantial
computational and storage resources for analysis and archival, which
are now available.  For our own use in the early stages of
the survey as well as for long-term multiwavelength studies, 
we will archive both the raw data and data
products from the data processing pipeline.

Large-scale pulsar surveys using the ALFA system have been
organized through a Pulsar ALFA (PALFA) Consortium of which the
present authors and others are members. The planning and execution of
PALFA surveys is a joint effort between NAIC and the Consortium to
obtain legacy results for use by the broader astrophysical community.  Similar
consortia have been organized for other Galactic science and for
surveys of extragalactic hydrogen.

The plan for the rest of this paper is as follows. Following a brief
description of the ALFA system in \S~\ref{sec:alfa}, in
\S~\ref{sec:survey} we describe the technical details and logistics of
our survey, including sky coverage, data acquisition and
processing, sensitivities, and archival of raw data and data products.
In \S~\ref{sec:results} we report on initial results from preliminary
survey observations that have so far resulted in the discovery of 11
new pulsars. Finally, in \S~\ref{sec:future}, we
outline our future plans and expectations for PALFA surveys.

\section{The ALFA System}\label{sec:alfa}

The ALFA feed horns are arranged in a close packed hexagon surrounding
a central horn at the Gregorian focus of the Arecibo
telescope. 
Orthomode transducers provide dual linearly-polarized
signals to cooled receivers.  The beams from the 
seven feeds are elliptically shaped
with equivalent circular beam sizes (FWHM) of 3.35~arcmin. 
Beam centers of the outer six beams fall on an ellipse of
size 11.0~arcmin $\times$ 12.8~arcmin. 
Efficient coverage of the sky requires that we compensate for
parallactic rotation of the beam pattern on the sky as the 
telescope azimuth changes. 
ALFA can be rotated relative to the telescope's azimuth
arm to accomplish this.  We note that, because the seven-beam
pattern is elliptical, there are small offsets of the beams from
their ideal positions as the feed is rotated.

\begin{figure}[!t]
\includegraphics[scale=0.45,angle=0]{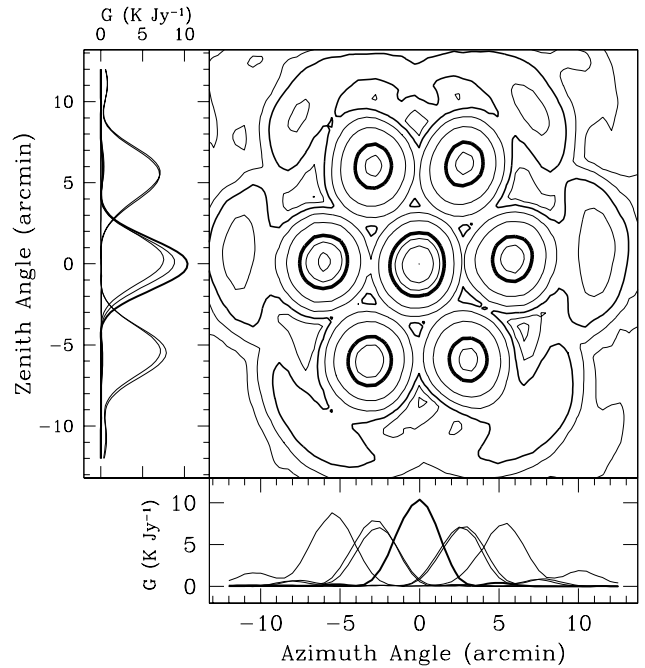}
\vspace {-0.9in}
\caption{\label{fig:alfabeams}
Contours of the telescope gain with the ALFA system averaged over
its passband (1225--1525~MHz).  Contours show
the maximum gain for a given azimuth and zenith angle from any one of
the seven beams and are for the average of the two polarization channels.
The gain values were measured on MJD 53129 at low zenith angles
using the extragalactic source 3C286 and
an assumed source flux density of 14.45~Jy for our measured band.  
Coutour levels are at
--1, --2, --3, --6, --9, --12, --15 and --19 dB from the central peak.
The heavy contour is at the --3 dB level and the next heaviest contour
outside the 6-beam pattern is for --12 dB.  Slices through the
centers of the individual beam patterns of the seven feeds 
are also shown at constant azimuth and constant zenith angles.
The equivalent circular beam width  (FWHM), averaged over all beams,
 is 3.35 arcmin at 1.42 GHz \citep{h04}.
}
\end{figure}

Figure~\ref{fig:alfabeams} shows
the measured gain contours for the feed systems.  
The on-axis gain is approximately
10.4~K~Jy$^{-1}$ at low zenith angles for the central beam but is
reduced to an average $\sim 8.2$~K~Jy$^{-1}$ for the other six
beams.
The system
temperature looking out of the Galactic plane $\sim 24$~K.  
Updated estimates of these system parameters and details on
their azimuth and zenith-angle dependences, 
are available at \url{http://alfa.naic.edu/performance}. 
Receiver signals are transported via optical fiber to intermediate-frequency
electronics and backend spectrometers in the control building.  

\begin{center}
\begin{figure*}[!ht]
\includegraphics[scale=0.8,angle=0]{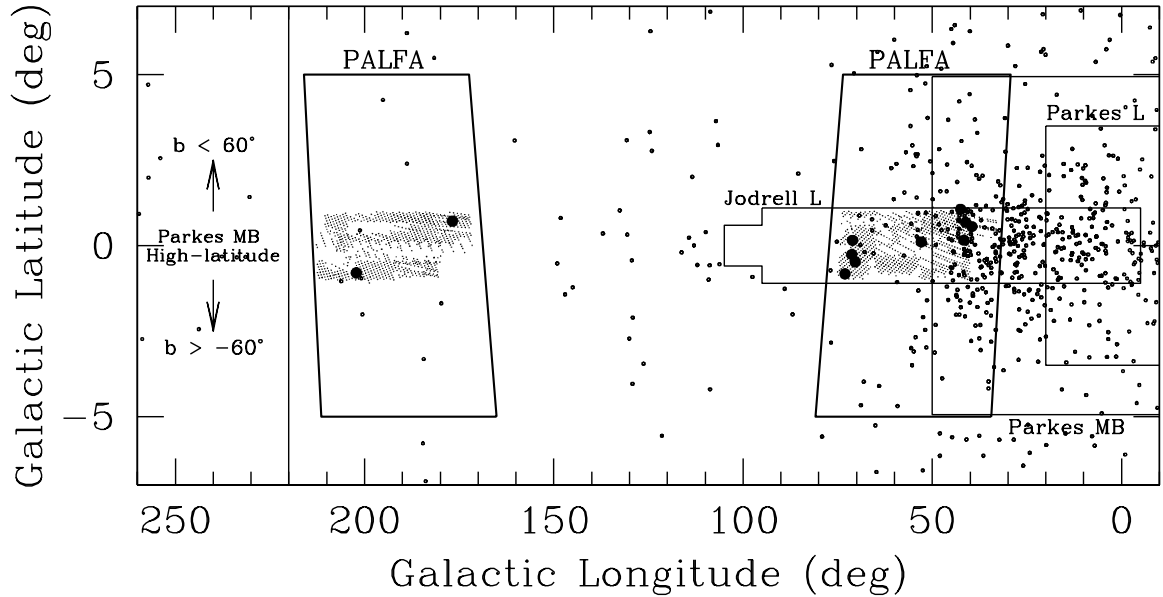}
\vspace {-1.5in}
\caption{ \label{fig:lbregions} 
Regions of the Galactic plane to be surveyed with PALFA, taking into
account declination limits of Arecibo and restricted to $\vert b \vert
\le 5^{\circ}$.  Hatched areas indicate regions covered so far in the
precursor survey and large filled circles represent newly discovered pulsars.
Small dots designate known pulsars.
We also show boundaries of several L-band surveys that have been made
in or near these regions, including the PMB survey and single-pixel
surveys with Parkes \citep{jlm+92} and Jodrell Bank \citep{clj+92}.
The Swinburne Intermediate Latitude Survey \citep{e+01}
covered the same longitude range as the PMB but at
latitudes $5^{\circ} \le \vert b \vert \le 15^{\circ}$. 
Arecibo surveys at 0.43~GHz have covered some of our proposed search
areas, but to distances much smaller than we can reach owing to the
limiting effects of interstellar dispersion and scattering.
}
\end{figure*}
\end{center}

\vspace{-0.4in}
Currently, we are using four Wideband Arecibo
Pulsar Processor (WAPP) systems \citep{dsh00} to process 100~MHz
passbands centered on 1.42~GHz for each ALFA beam.  As used in our
survey, the WAPPs compute 256 lags of the autocorrelation function
for each of two 3-level quantized polarization channels; correlation
functions for the two channels are summed before recording to disk as
2-byte integers at 64~$\mu$s intervals.

Within one year, we anticipate using new spectrometers that will process
the full 300 MHz bandwidth of the ALFA \hbox{frontend} system with 1024
spectral channels. The PALFA spectrometers will employ many-bit polyphase
filters implemented on field programmable gate array chips to provide
the channelization. We expect that mitigation of radio-frequency
interference (RFI) will be more robust with the new spectrometer
compared to the WAPP's 3-level  correlation approach. RFI shows 
a rich diversity
in the overall ALFA band. For the initial portion of the survey using
the WAPPs, we have therefore selected the cleanest 100~MHz portion of the
available spectrum, centered at a frequency of 1.42~GHz.

\vspace{0.2in}

\section{ALFA Pulsar Surveys}\label{sec:survey}
Previous surveys of the Galactic plane accessible with the Arecibo
telescope have either been conducted at the lower radio frequency of
0.43~GHz \citep{ht75b,sstd86,nft95} or with less sensitive systems at
1.4~GHz \citep{clj+92,m+01}. Our new survey with the ALFA system
promises to probe the pulsar population in the Arecibo sky
significantly more deeply than the previous surveys.

\subsection{Sky Coverage} \label{sec:coverage}

Our long term plan is to conduct comprehensive pulsar surveys of most of
the sky accessible with the Arecibo telescope (declinations between
$-1^{\circ}$ and $+38^{\circ}$) but with emphasis on the Galactic
plane (e.g., $\vert b \vert \lesssim 5^{\circ}$).  
The survey results discussed here are for
the inner Galaxy ($40^\circ \la
\ell \la 75^\circ$, $\vert b \vert \le 1^\circ$) and anti-center
($170^\circ \la \ell \la 210^\circ$, $\vert b \vert \le 1^\circ$)
regions. 
Figure~\ref{fig:lbregions} shows
the regions that have been and will be covered close to the Galactic plane.
The PMB pulsar survey covered the region $260^\circ \le \ell
\le 50^\circ$ and $\vert b \vert \le 5^{\circ}$, i.e., there is
some area of overlap in the inner Galaxy.  
Later we will conduct a survey at 
intermediate latitudes up to $\vert b\vert \sim 20^{\circ}$ to optimize the
search for relativistic binary systems and MSPs.

\begin{center}
\begin{figure}[!ht]
\includegraphics[scale=0.43,angle=0]{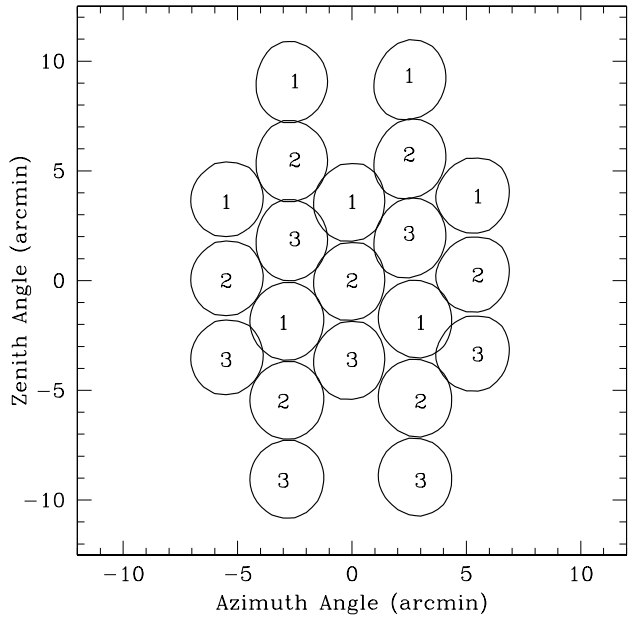}
\vspace {-0.9in}
\caption{ \label{fig:dense} 
ALFA beam locations on the plane of the sky showing the half-power beam
widths for three pointings derived from the data presented in
Figure~\ref{fig:alfabeams}. The pointings are labeled 1, 2 and 3 in a
dense sampling grid that covers nearly all of the solid angle with at
least half the gain of the relevant beam.  Sparse sampling consists of
making only one of these pointings.  Note the ellipticity of the beams
and their pattern for a given pointing.
}
\end{figure}
\end{center}

Our strategy for sampling the sky employs two methods for 
maximizing the efficiency and sensitivity of the survey.  
As shown in Figure~\ref{fig:dense}, 
three adjacent pointings of ALFA are required
to tile the sky with gain equal to or greater than half the maximum gain. 
Rather than using this dense sampling scheme, we have so far
adopted a sparse sampling scheme \citep{fre03} that makes only one 
out of three of these pointings.
Monte Carlo simulations \citep{vc04,fg04}
indicate that sparse sampling should detect $\sim 2/3$ of the pulsars in the
surveyed region.  
This scheme has the advantage that more solid angle is covered per unit time,
though much of it at substantially less than half the full gain. 
The sparse sampling approach exploits the large sidelobes
for the outer six beams, which are $\sim 16$\% ($-$8 dB) of the peak
gain centered $\sim 5$ arcmin from the beam axis 
(see Figure~\ref{fig:alfabeams}).
 The gains of these
sidelobes are approximately 0.7 and 1.6 times the on-axis gains for
the Green Bank Telescope and the Parkes 64-m telescope, respectively, 
and thus provide significant sensitivity.

\vspace{-0.365in}
Later on, we will make the two additional passes
needed to achieve dense coverage.  Despite the smaller numbers of new
pulsars expected in these subsequent passes, they will yield more pulsar
discoveries than if we were to extend the sparse coverage to higher
Galactic latitudes, where the pulsar density decreases rapidly.





We are also exploring a multiple-pass strategy, where a given sky
position is observed two or more times. This approach is motivated by
the fact that pulsar flux densities are highly variable (over and
above the fundamental pulsation), sometimes by more than an order of
magnitude, due to a number of intrinsic and extrinsic causes,
including nulling and mode changes common to many pulsars, eclipses,
and interstellar scintillation (both long-time-scale
refractive scintillation and fast, diffractive scintillation). RFI is
also episodic. Our simulations suggest that pulsars can be missed in
single-pass strategies but that any improvement from multiple pass
approaches depends on the details and prevalence of flux
modulations. 
We are in the process of comparing 
different strategies while also using simulations to fully optimize
our usage of ALFA.

\subsection{Data Analysis }

To maximize the pulsar yield and overall science return from the PALFA
survey, we are processing the data twice.  During the
observations, incoming data are transferred to the Arecibo Signal
Processor\footnote{\url{http://astron.berkeley.edu/$\sim$dbacker/asp.html}}, 
a computer cluster that processes the data in quasi-realtime after
reducing the time and frequency resolution to increase throughput.  
This ``quicklook'' analysis,
described below, is primarily sensitive to pulsars with $P
\gtrsim 30$~ms, which are expected to make up the bulk of all
discoveries. We are currently developing an offline data analysis
scheme which retains the full resolution of the data and will be
sensitive to MSPs and pulsars in short period binary systems as well 
as to pulsars with large values of dispersion measure (DM). 
In addition to using a number of different pulsar search codes and
algorithms, this latter analysis pipeline will also take advantage of
the multiple beam data acquisition for RFI excision.  Analysis of the
raw data will be done on several computer clusters at the home
institutions of members of the PALFA Consortium.  Further details will
be published elsewhere.

The quicklook pipeline uses freely available pulsar data analysis
tools \citep{lor01} to unpack and transform the correlation functions
from the WAPPs to spectra with 256 channels every 64 $\mu$s. 
The data are decimated in frequency and time
 by factors of 8 and 16, respectively, to allow quasi-realtime processing. 
The resulting data sets with
32 frequency channels and 1024 $\mu$s time resolution are then
corrected for the effects of interstellar dispersion by appropriately
delaying low frequency channels relative to the highest one. This
process is carried out for 96 trial values of DM
in the range 0--980~pc~cm$^{-3}$. The step size in DM is
approximately 2, 4, 8, 16 and 32 pc~cm$^{-3}$ with changes in step
size at approximately 62, 124, 253, and 506 pc~cm$^{-3}$. 
Two different searches are
then carried out on the resulting dedispersed time series, one for
periodic signals and a second for isolated pulses.

\begin{center}
\begin{figure}[ht]
\includegraphics[scale=0.49,angle=0]{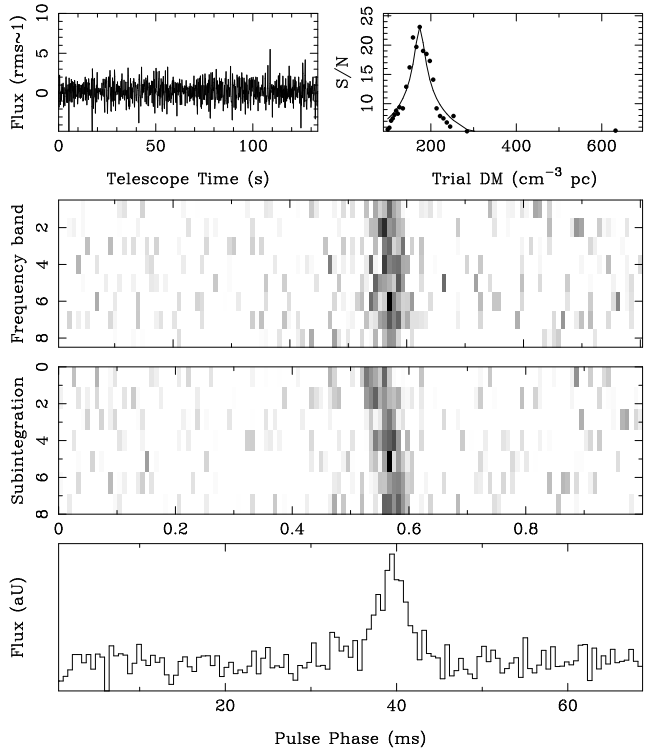}
\caption{\label{fig:seek}
Sample periodicity search output from the quicklook analysis showing
the discovery of the 69-ms pulsar \psrone.  Top left: a coarse version
of the dedispersed time series used to assess basic data quality. Top
right: S/N as a function of trial DM. The points show detections in
the periodicity search while the curve is the theoretically expected
response given the system parameters and pulse width.  Middle panels:
gray scales showing pulse intensity as a function of sky frequency
(quantized into frequency bands) and observing time (quantized into
subintegrations). Bottom: folded pulse profile obtained by integrating
over the whole pass band. 
}
\end{figure}
\end{center}

The analysis for periodic, dispersed pulses
follows most standard pulsar search schemes (see,
e.g., Lorimer \& Kramer 2005)\nocite{lk05}, and the software used for
this analysis is based on code developed for an earlier survey
\citep{lkm+00}.  In essence, the procedure is to look for harmonically
related signals in the amplitude spectrum (the magnitude of
the Fourier transform)
 of each dedispersed time series.  To increase sensitivity to
signals with narrow duty cycles, which have  many harmonics in the Fourier
domain, the amplitude spectra are incoherently summed so that up to
2, 4, 8 and 16 harmonics are combined. A list of candidates with
signal-to-noise (S/N) ratios above 8 is then formed and the data are
folded in the time domain to produce a set of diagnostic plots of the
form shown in Figure~\ref{fig:seek}.  To simplify the viewing of the
search output, a web-based browsing system was developed 
for examination of 
candidate signals during an observing session.  The most
promising pulsar candidates are filed for future observation and follow up. 

\begin{figure}[ht]
\includegraphics[scale=0.45,angle=0]{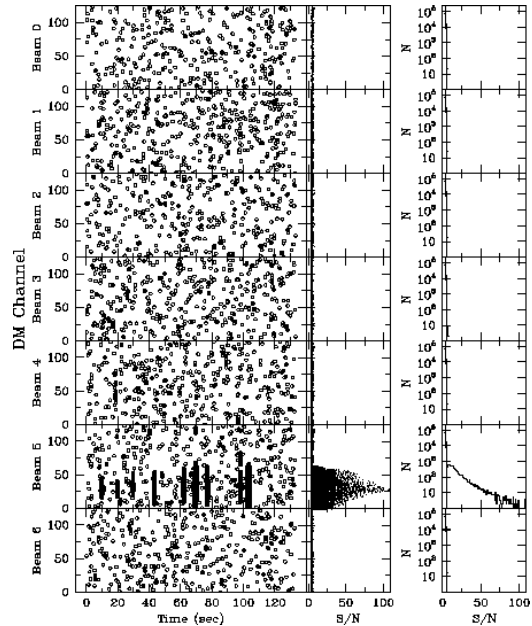}
\vspace {-0.0in}
\caption{\label{fig:spout}
Single-pulse search output for 
a follow-up observation of the 1.2-s pulsar J0628+09. Each row
shows data collected by one of the seven beams during the
pointing. From left to right the plots show: 
scatter plot of events with S/N $> 5$ vs. time and 
dispersion-measure (DM) channel; 
scatter plot of DM channel and S/N for events;
and the number of pulses versus S/N. Individual pulses from J0628+09
are clearly seen only in beam 5.   The 
distribution of events vs. DM depends on the pulse shape and width
\citep{cm03}.
}
\end{figure}

In parallel with the periodicity analysis, we also search for isolated
pulses in the 96 dedispersed time series based on code developed by
\cite{cm03}. In brief, threshold tests are made on each dedispersed
time series after smoothing it by different amounts to approximate
matched-filter detection of pulses with different widths. 
In addition, we consider events defined
by clusters of above-threshold samples in a ``friends-of-friends''
algorithm \citep{f95}.  For each pointing, diagnostic plots similar to 
those shown in Figure~\ref{fig:spout} are generated for inspection
within the candidate browsing system. This example shows data for the
1.2~s pulsar \psrtwo, which we discovered in our single-pulse search
through its bright individual pulses (see \S \ref{sec:psrtwo}).
The three panels for each of the seven ALFA beams show, from left
to right:  events above a threshold S/N $> 5$ vs. time and DM channel; 
a scattering plot of DM channel vs. S/N; and a histogram of S/N.
Events appear with largest S/N in the DM channel that best matches
the pulsar's DM; they also appear in neighboring DM channels with lower
S/N that depends on the pulse width \citep{cm03}.

\vspace{-0.4in}
\subsection{Database Management and Archiving}

PALFA survey results are archived in a
MySQL\footnote{http://www.mysql.com} database system, which stores sky
coverage and data quality information along with results from the data
processing. This system is also designed to record 
the results from several different
search codes that implement the processing system described above to
allow  comparison and optimization.  
The MySQL database includes
an observational table with 
fields that characterize specific telescope pointings, the
resulting raw data files, and ancillary information about the
observations.  Another table reports the results of the preliminary,
quicklook data analysis, with fields that describe candidate signals and whether
they correspond to known pulsars or not.  
There is another table that tracks the content
and location of portable disk drives used to transport data from the
observatory to processing sites.  Additional tables report results from
the data processing that uses the full data resolution,  a list of refined
pulsar candidates, and the status of confirming and other follow-up 
observations.

A data archive is under development at
the Cornell Theory Center and will include the 
original raw data as well as analysis products and database mining tools, 
accessible through a data gateway.\footnote{http://arecibo.tc.cornell.edu}

\subsection{Search Sensitivity}

Data taken in our preliminary survey dwell on particular sky positions
for 134~s for the inner Galaxy and 67~s for anticenter directions.
The minimum detectable flux density $\Smin$ for PALFA with these
parameters is a factor 1.6 smaller than for the PMB survey, implying a
maximum distance $\Dmax \propto \Smin^{-1/2}$ about 1.3 times larger
for long-period pulsars.  The sampled volume on axis is accordingly
about a factor of two larger for long-period pulsars. 
In our full-resolution analysis, 
the volume increase is even larger for small periods
 owing to the smaller PALFA channel
widths and the shorter sample interval.  For $P\lesssim 10$ ms, the
searched volume increase can be a factor of 10 or more.

\begin{center}
\begin{figure}[ht]
\includegraphics[scale=0.40,angle=0]{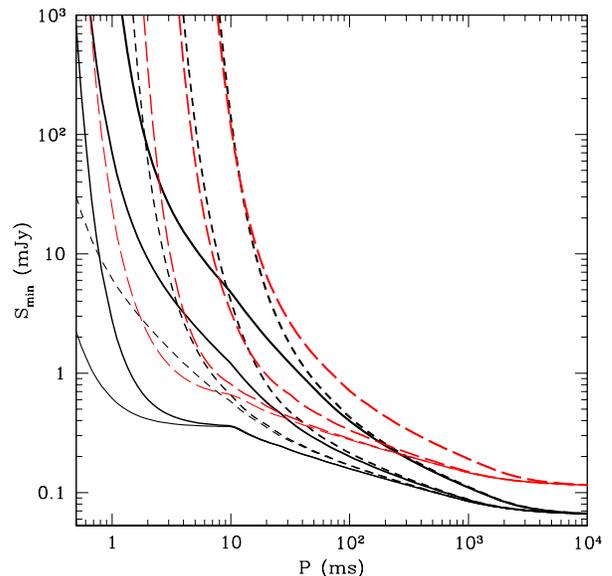}
\vspace {-0.9in}
\caption{\label{fig:smin}
Theoretical minimum detectable flux density ($\Smin$) vs.  $P$ for
different values of DM.  
%
Short dashed lines: for the coarse-resolution PALFA data analyzed 
with the quicklook software that led to the discoveries reported in
this paper.   
Long dashed lines: for the Parkes
multibeam survey, which used 96 channels across 288 MHz and 250 $\mu$s
sampling for scan durations of 2100 s.  
Solid lines: for full-resolution PALFA data.
For each set of curves, DM
values from the lowest to the highest curve are 1, 200, 500
and 1000~pc~cm$^{-3}$. The breakpoint at $P \sim 10$ ms for the solid
curves occurs because we assume that the intrinsic pulse duty cycle
scales as $P^{-1/2}$ with a maximum of 0.3, which occurs at this
period. Above 10 ms, the number of harmonics contributing to
detections increases from 1 to 16 (the maximum searched) as the duty
cycle gets smaller.  A threshold of 10$\sigma$ is used.
}
\end{figure}
\end{center}

\vspace{-0.4in}

Figure~\ref{fig:smin} shows idealized plots of $\Smin$ versus $P$ for
four values of DM (from 1 to $10^3$~pc~cm$^{-3}$) using post-detection
dedispersion followed by a standard Fourier analysis with harmonic
summing. 
The values of $\Smin$ include the effects of radiometer noise
and pulse smearing from instrumental effects combined with dispersion and
scattering in the interstellar medium using the ``NE2001'' electron
density model \citep{cl02}.   Scattering has been calculated in the
model for the particular direction $b = 0^{\circ}$ and
$\ell = 40^{\circ}$.   Directions at higher latitudes will show less
scattering and better sensitivity for large values of DM.  
The results are also based on
the assumption that pulse amplitudes are constant over the observation
time that spans many pulse periods, which is obviously an idealization.  
Our calculations for the
PMB survey do not include high-pass filtering in both hardware and software
that degrades the sensitivity to long-period pulsars
\citep[see][]{m+01}.
No high-pass filtering is done in our analyis, either in hardware
or in software.
We emphasize that the curves in Figure~\ref{fig:smin} should be
interpreted as {\it lower bounds} on the true values of $\Smin$
because real-world effects such as RFI and receiver gain variations
will raise the effective threshold of the survey.  Our 
quicklook analysis described above, which analyzes data after
decimation in time and frequency, has detection curves
about 60\% more sensitive than those for the PMB survey except for
$P\lesssim 10$ ms, for which the large sampling time of the quicklook
analysis significantly degrades the sensitivity.

\section{Initial Results}\label{sec:results}

\begin{figure}[ht]
\includegraphics[scale=0.43,angle=0]{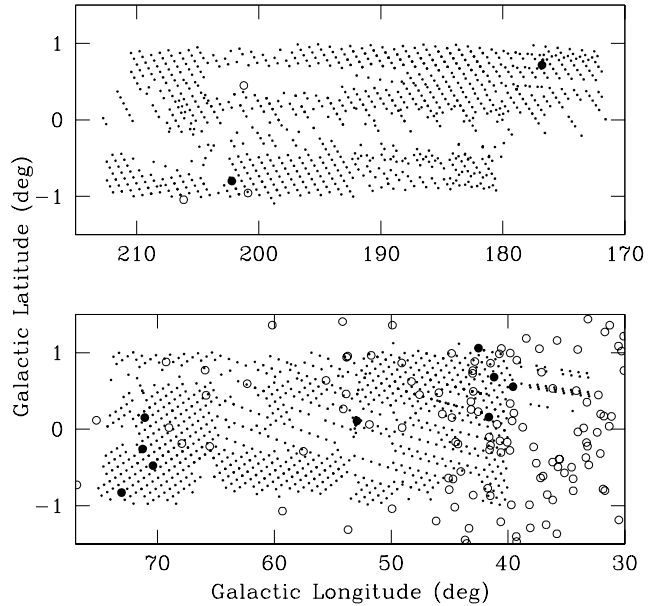}
\vspace {-0.9in}
\caption{ \label{fig:lbzoom} 
Regions of the Galactic plane surveyed with PALFA to date showing the
Galactic anti-center region (top) and inner Galactic plane
(below). Dots denote the pointing centers of each 7-beam cluster, filled
circles show newly discovered pulsars while the open circles designate
previously known pulsars.
}
\end{figure}

For our preliminary survey observations carried out between 2004
August and 2004 October, we have used 17.1 hours of telescope time for
919 pointings in the Galactic anti-center and 32.2 hours for 865
pointings in the inner Galaxy, covering 15.8~deg$^2$ and 14.8~deg$^2$
in each region respectively.  These numbers and their graphical
presentation in  Figure~\ref{fig:lbzoom} were obtained using the
MySQL database described earlier.
Table~\ref{tab:precursor.new.pulsars} lists the 11 new pulsars that we
have found so far. Table~\ref{tab:precursor.known.pulsars} 
lists the detection statistics of 29 previously known pulsars also seen
in the quicklook analysis pipeline. Not included here is a 
detection of the 1.55-ms pulsar~B1937+21, which was undetected
due to the coarse time resolution of the quicklook pipeline. The
pulsar was, however, easily detected when the raw data were folded at
their full resolution. 
The high time-resolution data pipeline mentioned above 
will allow detection of any MSPs missed in the quicklook analysis.

\subsection{General Remarks}

Four of the pulsars discovered in the inner Galaxy (J2009+33,
J2010+32, J2011+33 and J2018+34) are in the northernmost region
detectable from Arecibo; all of these have DMs between 220 and
350~pc~cm$^{-3}\,$. Some, if not all, of these objects are possibly
associated with the Cygnus region, where a large number of supernovae
will have produced many relatively young pulsars
(e.g., Vlemmings et al.~2004\nocite{vcc04} and references
therein). Another four objects (J1901+06, J1904+07, J1905+09 and
J1906+07) are in the southernmost region visible from Arecibo, where
the density of pulsars is known to be higher. This region was
previously covered by the PMB survey, suggesting that our
precursor survey indeed already surpasses the depth of the PMB survey
when a conventional pulsar search analysis is done.  Subsequent to our
discovery, J1906+07 was identified in the acceleration search output
of the PMB data (Lorimer et al. 2005, in preparation). 

All previously known pulsars were detected in our pointings if they
were within one beam radius of one of the ALFA beams.
In addition, we detect some strong pulsars
several beam radii from the nearest beam center.  A coarse analysis
suggests that our detection rate is consistent with what we expect
from the sparse-sampling strategy discussed earlier 
based on simulations.  A detailed analysis will be done as we
continue the survey.

The single-pulse search analysis is notably successful in detecting
6 out of 11 of the new pulsars and 21 out of 29 of the known pulsars.
These statistics are consistent with the fact that the known pulsars
tend to be stronger than our new detections.   The single-pulse analysis
is valuable both for  corroborating candidate detections
from the periodicity analysis and, as  we have shown in
the case of J0628+09, for identifying pulsars that are missed
in the periodicity search owing to the intermittency of their pulses. 

\subsection{\psrone}

The first pulsar discovered in the ALFA survey, \psrone, also has the
shortest period among the pulsars discovered so far, $P = 68.7$~ms.  
While time for follow-up
observations on this and the other pulsars discovered has so far been
limited, we have made some multi-frequency and timing observations
of \psrone.  Using the 
{\sc tempo}\footnote{http://pulsar.princeton.edu/tempo} software package to
analyze 83 arrival times from \psrone\ spanning a 257-day baseline, 
we obtain results presented in Table~\ref{tab:parms}. The timing
model implies that \psrone\ is a young isolated pulsar with a
characteristic age $\tau_c = P/2\dot P =  82\,$kyr, 
a surface magnetic field strength 
$B = 3.2\times 10^{19}\sqrt{P\dot P}~G =  
9.6 \times 10^{11}$~G (assuming a dipolar field) 
and spin-down energy loss rate $\dot{E}= I\Omega\dot\Omega = 1.6
\times 10^{36}I_{45}$~erg~s$^{-1}$ (where $\Omega=2\pi P^{-1}$ and 
$I_{45}$ is the moment of inertia in units of $10^{45}$ g cm$^2$).

\begin{figure}[!t]
\includegraphics[scale=0.40,angle=0]{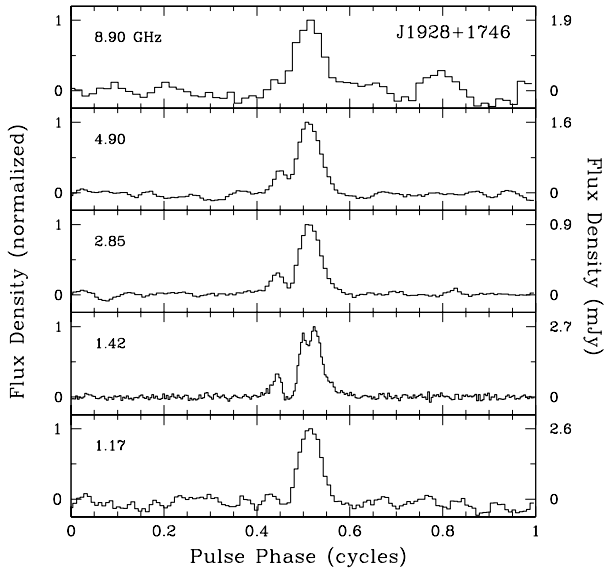}
\vspace {-0.9in}
\caption{ \label{fig:J1928+1746profiles} 
Pulse profiles for \psrone\ at five frequencies from 1.2 to 
8.9 GHz obtained with integration times of 135\,s, 4173\,s, 804\,s, 900\,s 
and 106\,s, from low to high frequency.  The flux density scale 
on the right-hand side is accurate to approximately 20\%.
}
\end{figure}

Multifrequency observations from 1.1 to 9 GHz shown in
Figure~\ref{fig:J1928+1746profiles} suggest that the radio spectrum is
nearly flat, $S_{\nu} \propto\nu^{+0.2\pm 0.3}$. The quoted error
reflects empirical departures from the fit and thus includes
any systematic calibration errors or random errors from scintillations.  
Estimates of the flux densities are coarse because we have simply scaled
the signal-to-noise ratios of the average pulse amplitudes and used typical
values for the gain and system temperature.  
The flux densities  at the higher two frequencies 
are likely to be influenced by modulations from
interstellar scintillation (based on DM and the likely distance). 
High-frequency surveys are naturally biased toward the discovery of
objects with flatter spectra than surveys at lower frequencies.  Also,
young pulsars appear to have flatter spectra \citep{l+95}, so
high-frequency surveys of the Galactic plane will be less biased against
them. \psrone\ appears to be a prototype flat-spectrum object
of which we can expect to find more in our survey.

\begin{figure}[!t]
\includegraphics[scale=0.65,angle=0]{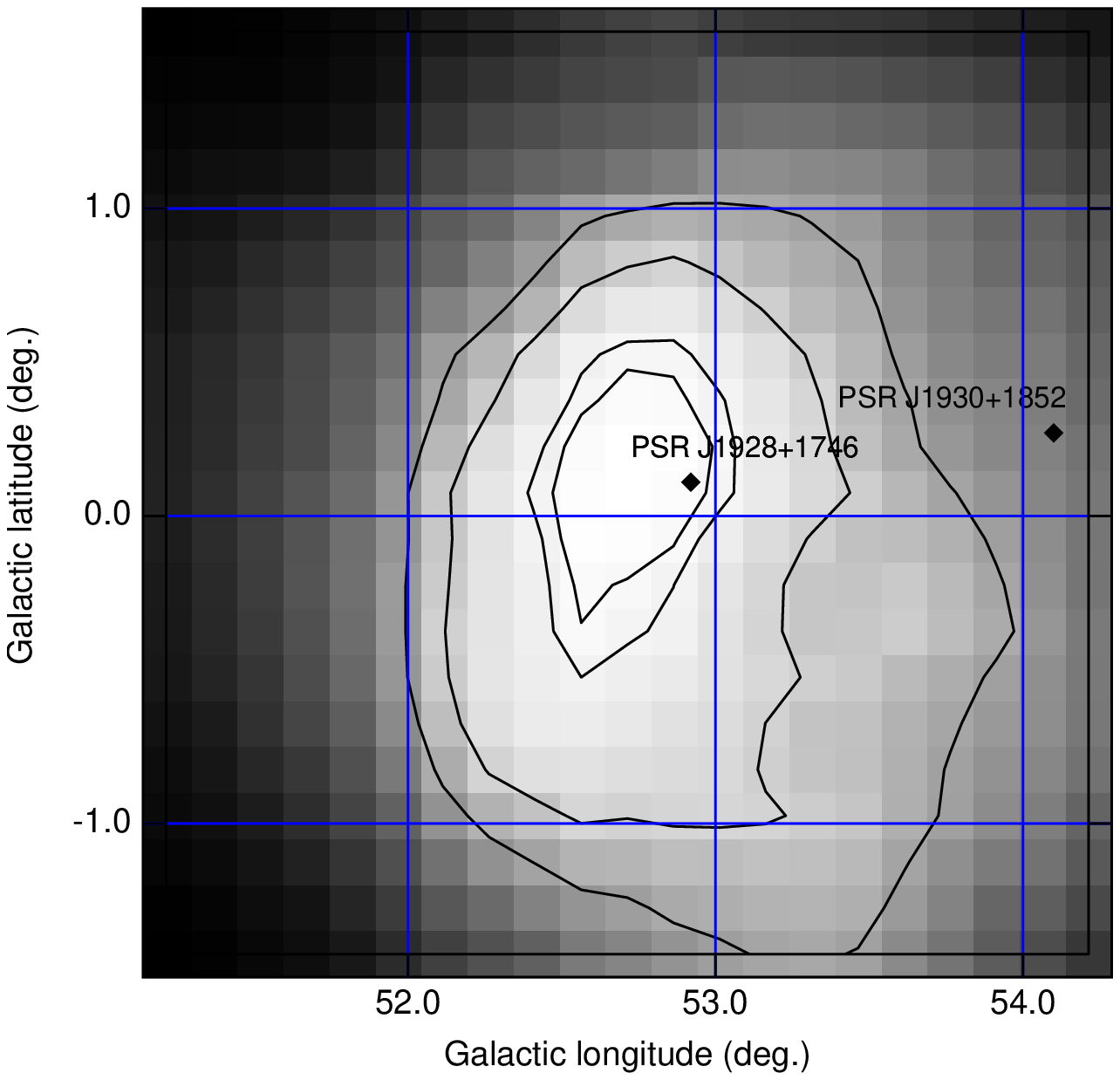}
\caption{ \label{fig:3EGJ1928}
EGRET localization test statistic of the high-energy gamma-ray source
3EG~J1928+1733 \citep{hbb+99}. Contours delimit probability regions,
from innermost to outermost, of 50\%, 68\%, 95\%, and 99\% for the
location of the gamma-ray emission. In addition to \psrone\ which lies
close to the center, we also show the next nearest pulsar, J1930+1852,
previously advanced \citep{clb+02} as a possible counterpart
to 3EG~J1928+1733 even though it lies outside the 99\% contour.
}
\end{figure}

As shown in Figure~\ref{fig:3EGJ1928}, \psrone\ lies well within the 
localization map for the unidentified EGRET source 3EG~J1928+1733. 
The EGRET source shows significant variability \citep{t+01} that is
indicative of a blazar, but has a photon index, $\Gamma=2.23\pm0.32$, not
inconsistent with those of known pulsars.
If \psrone\ is the radio pulsar counterpart to 3EG~J1928+1733, the
implied efficiency for conversion of spin-down energy into gamma-rays
is $\eta_\gamma \equiv L_\gamma/\dot E = 22\% \Omega_{\gamma} (d /6 \,{\rm
kpc})^2$, where $\Omega_{\gamma}$ is the solid angle (sr) swept
out by the pulsar's beam
 and a photon index of $-2$ is assumed for the gamma-ray spectrum.  
While the nominal efficiency is higher than that of
any of the confirmed gamma-ray pulsars \citep{tbb+99}, we note that the
above calculation is strongly dependent on the uncertain beaming fraction
and on the  DM-derived distance to \psrone\ of 6 kpc. 
Also, the flux measurement used to calculate the efficiency from the
3EG catalog \citep{hbb+99} is the largest (and most significant) value,
so the implied efficiency should be viewed as an upper bound.
Two other young pulsars recently discovered within EGRET error boxes,
J2021+3651 \citep{rhr+02} and J2229+6114 \citep{hcg+01} have
similarly high inferred efficiencies.  These pulsars will be excellent
future targets for the Gamma-ray Large Area Space Telescope (GLAST).

\subsection{\psrthree}

\psrthree\ was initially attributed a spin period half of its actual
value owing to the presence of its interpulse.  
Very recently, this object has been
found to be a binary pulsar by comparing the parameters in 
Table~\ref{tab:precursor.new.pulsars} with entries in the PMB
database and by making new observations at Jodrell Bank.  The orbital
period is 3.98 hr, the eccentricity is 0.085 and its
projected orbital semi-major axis is $a_1\sin i = 1.42$ light s,
yielding a mass function  of 0.11 $M_{\odot}$.   Discovery of the binary
nature of this source and a discussion of its properties may
be found in Lorimer et al. (2005, in preparation). 

The discovery of \psrthree\ underscores the power and great potential
of PALFA surveys for finding binary pulsars. The large sensitivity of
the Arecibo telescope allows us to use integration times short enough
that many binaries will be detected without recourse to acceleration
searches, as has been necessary with the PMB survey.   Acceleration
searches lead to a greater number of statistical trials which, in turn,
require  a higher detection threshold to minimize the number of
``false-alarm'' detections.

\subsection{\psrtwo}
\label{sec:psrtwo}

Many pulsars show large modulations of the pulsed flux density.  In
some cases, it is easier to detect single pulses than the
periodicity in a Fourier analysis \citep{n99, mc03}. In our quicklook
analysis, most of the known pulsars and three of the pulsars
discovered in the periodicity search (J0540+32, J1904+07 and J2011+33)
appear strongly in the single pulse search and less so for three other 
pulsars.  Some single pulses of
PSR~J1904+07 are detected with S/N $\sim 36$, more than double that
found by the periodicity search. 
\psrtwo \ was discovered {\em only} by detection of its very sporadic 
single pulses, some of which have peak 
S/N$\sim 100$. 
The discovery data set had
only three large pulses in a 67 s scan, too few to allow the pulsar's 
detection in the periodicity search.  A periodicity of 2.48 s was
determined from the arrival times of those pulses.  
Subsequent observations with a greater number of strong pulses
and an above-threshold detection in the periodicity analysis
have allowed us to determine the true period of 1.24 s.

The discovery of \psrtwo\ clearly demonstrates the importance of
single pulse searches. As shown in Figure~\ref{fig:spout}, these
searches are enabled by the simultaneous measurements in
multiple beams, which allow discrimination
between RFI and celestial events.  Extrapolating from the present
sample to the whole survey, we can expect to find 
a significant number of pulsars through their single-pulses and not
through their periodicity.
The analysis also may detect 
radio transients from non-pulsar objects,
a plausible outcome given 
the recent discovery of a transient radio
source in the direction of the Galactic center \citep{hlk+05} and a
number of other transient radio sources found in a single-pulse
analysis of the PMB survey (McLaughlin et al. 2005, in preparation). 

\section{Future Plans and  Expectations}\label{sec:future}

We have described the initial stages of a large-scale survey for
pulsars using ALFA, the seven-beam system at 
the Arecibo Observatory that operates at 1.4 GHz. 
Our discovery of 11 pulsars from precursor
observations --- using a preliminary data acquisition system that sampled only
1/3 of the available bandwidth followed by a quick-look analysis --- 
is extremely encouraging. 
A new spectrometer that uses the full bandwidth will become available
within the next year.
The full data processing pipeline,
now under development, will have excellent sensitivity to MSPs and is
expected to yield further pulsar discoveries in our existing data.
This pipeline will include masking of RFI in the frequency-time plane
prior to dedispersion, a new matched-filtering search algorithm  for
events that have a broader range of frequency-time signatures than
those encountered for pulsars, and compensation for 
acceleration in binary systems.  

In the near future, we expect to begin  regular timing programs 
on several telescopes  to obtain precise determinations of the spin 
and astrometric parameters of these pulsars and others that will 
be discovered. 

The full survey will take more than five years,
depending largely on allocation of telescope time. 
Numerical models of the pulsar
population, calibrated by results from the PMB survey and incorporating
measured characteristics of the ALFA system, suggest that as many as
1000 new pulsars will be discovered.

The raw data from the search as well as the data products
from the search analysis will be archived and made available to
the broader community via a web-based portal.   Initially,
the database system will enable our own mining of the data for
new pulsars and perhaps other astrophysical signals.  Later, we 
expect the system to provide opportunities for multi-wavelength
searches, such as identification of radio counterparts to 
X-ray sources or to 
candidate gamma-ray pulsars seen in data from GLAST. 

\acknowledgments

We are grateful to the staff at Arecibo, NAIC and ATNF for their hard work
on the ALFA frontend feeds and receivers and associated
backend digital spectrometers and recording systems.  In particular,
we thank Arun~Venkatarman, Jeff~Hagen, Bill~Sisk and Steve~Torchinsky 
at NAIC and Graham~Carrad at the ATNF. 
This work was supported by the National Science Foundation
through a cooperative agreement with Cornell University to operate the
Arecibo Observatory.  NSF also supported this research through grants
AST-02-06044 (UC Berkeley),
AST-02-05853 (Columbia University), 
AST-02-06035 (Cornell University), and 
AST-02-06205 (Princeton University).  
Pulsar research at UBC is supported by an NSERC Discovery Grant.
The Arecibo Signal Processor (ASP) is partially funded by an NSERC
RTI-1 grant to IHS.
ZA acknowledges support from grant NRA-99-01-LTSA-070 to NASA GSFC. 
NDRB is supported by a MIT-CfA Fellowship at Haystack Observatory.
DJC is funded by the Particle Physics and Astrophysics Research Council 
	in the UK. 
CAFG acknowledges support from the Natural Sciences and Engineering Research
	Council of Canada (NSERC) in the form of Undergraduate Student Research
	Awards (USRA).
JLH is supported by National Natural Science Foundation of China (10025313
and 10328305).
LK holds an NSERC PGS-M.
VMK is a Canada Research Chair and NSERC Steacie Fellow, and is
supported by NSERC, FQRNT, CIAR and the Canada Foundation for
Innovation.
DRL is a University Research Fellow funded by the Royal Society.  
IHS holds an NSERC UFA.
We thank Manuel Calimlim, Johannes Gehrke, David Lifka, Ruth Mitchell,
John Zollweg and the Cornell Theory Center for their work on
developing the survey database system, porting of code,
 and discussions about data mining.
Database work at Cornell is supported by 
NSF RI Grant 0403340, by a Microsoft E-Science
Grant, and by the Unisys Corporation. 
Any opinions, findings, conclusions or recommendations expressed in
this material are those of the authors and do not necessarily reflect the
views of the sponsors.

{ 
\begin{deluxetable}{lrrrrrll}
\tablewidth{500pt}
\tablecaption{\label{tab:precursor.new.pulsars}
Pulsars Discovered in the PALFA Precursor Survey}
\tablecolumns{8}
\tablehead{
PSR  & \colhead{RA (J2000)} & \colhead{DEC (J2000)} &
\colhead{$P$} &\colhead{$\widehat{\rm DM}$} &\colhead{$\langle S/N \rangle$} & SP? &
Comments
\\
        & & & (ms)  & (pc cm$^{-3}$) & & \\
}
\startdata
J0540+32        &
05$^{\rm h}$40$^{\rm m}$38$^{\rm s}$ & $+32^\circ$02\arcmin 19\arcsec &
  524  & 120   & 36 & Y &
Strong, sporadic single pulses \\
J0628+09        & 
06$^{\rm h}$28$^{\rm m}$49$^{\rm s}$ & $+09^\circ$09\arcmin 59\arcsec &
1241  &  88   & -- & Y &
Discovered as S/N=40 single pulses \\
J1901+06        &  
19$^{\rm h}$01$^{\rm m}$36$^{\rm s}$ & $+06^\circ$09\arcmin 36\arcsec &
832  & 162   & 14 & Y &
\\
J1904+07        &
19$^{\rm h}$04$^{\rm m}$09$^{\rm s}$ & $+07^\circ$39\arcmin 41\arcsec &
  209  & 275   & 15 & Y &
Strong, sporadic   \\
J1905+09        &
19$^{\rm h}$05$^{\rm m}$16$^{\rm s}$ & $+09^\circ$01\arcmin 22\arcsec &
  218  & 452   & 14 & N & \\
J1906+07        &
19$^{\rm h}$06$^{\rm m}$51$^{\rm s}$ & $+07^\circ$49\arcmin 01\arcsec &
  144  & 217   & 11 & N &
Interpulse; original detection\\
&&&&&&& at 72 ms; binary with $P_{\rm orb} = 3.98$ hr \\
J1928+1746      &
19$^{\rm h}$28$^{\rm m}$43$^{\rm s}$ & $+17^\circ$46\arcmin 23\arcsec &
   69  & 174   & 19 & N &
First ALFA pulsar; flat spectrum\\
J2009+33        &
20$^{\rm h}$09$^{\rm m}$39$^{\rm s}$ & $+33^\circ$25\arcmin 58\arcsec &
 1438  & 254   & 13 & N &
Sporadic\\
J2010+32        &
20$^{\rm h}$10$^{\rm m}$21$^{\rm s}$ & $+32^\circ$30\arcmin 22\arcsec &
 1442  & 350   & 23 & N &
\\
J2011+33        &
20$^{\rm h}$11$^{\rm m}$47$^{\rm s}$ & $+33^\circ$21\arcmin 49\arcsec &
  932  & 300   & 30 & Y &
Sporadic \\
J2018+34        &
20$^{\rm h}$18$^{\rm m}$54$^{\rm s}$ & $+34^\circ$32\arcmin 44\arcsec &
  387  & 226   & 24 & Y &
\enddata
\tablecomments{RA and DEC are the right ascension and
declination for the center of the beam where the pulsar was
found. Typical half-width uncertainty in pulsar position is one beam
radius (about 1.6 arcminutes) in both coordinates, except for \psrone.
$\widehat{\rm DM}$ is the DM value at which the
search algorithm identified the pulsar with maximum S/N.
$\langle S/N \rangle$  is the
signal-to-noise ratio of the averaged pulse shape.
SP denotes whether individual pulses from this pulsar
were detected in the single-pulse search.
}
\end{deluxetable}
}

\begin{deluxetable}{lrrrrrrr}
\tablewidth{350pt}
\tablecaption{\label{tab:precursor.known.pulsars} Previously Known Pulsars
Detected in the PALFA Precursor Survey }  
\tablecolumns{8}
\tablehead{
PSR  & \colhead{$P$} &  \colhead{DM} & \colhead{$\widehat{\rm DM}$}&
\colhead{S$_{1400}$} & \colhead{$\langle {\rm S/N} \rangle$} &
        \colhead{$\Delta\theta$} & SP? \\
        & (ms)  & \multicolumn{2}{c}{(pc cm$^{-3}$)} & (mJy) & & ($'$) & \\
}
\startdata
J0631+1036 &  287 &  125 & 148 & 0.8  &  76 & 6.7 &Y\\
J1855+0307 &  845 &  403 & 410 & 0.97 &  40 & 3.2 &Y\\
B1859+07   &  644 &  253 & 282 & 0.9  &  42 & 2.3 &Y\\
B1903+07   &  648 &  245 & 226 & 1.8  & 161 & 0.6 &Y\\
B1904+06   &  267 &  473 & 508 & 1.7  &  82 & 2.4 &Y\\\\
J1904+0800 &  263 &  439 & 424 & 0.36 &  17 & 2.0 &N\\
J1905+0616 &  990 &  258 & 283 & 0.5  &  41 & 1.8 &Y\\
J1906+0912 &  775 &  265 & 240 & 0.32 &  12 & 5.4 &Y\\
J1907+0740 &  557 &  332 & 353 & 0.41 &  20 & 2.3 &Y\\
J1907+0918 &  226 &  358 & 353 & 0.29 &  18 & 4.7 &N\\   \\ 
B1907+10   &  284 &  150 & 198 & 1.9  &  57 & 1.9 &Y\\
J1908+0734 &  212 &   11 & 46  & 0.54 &  13 & 1.1 &Y\\
J1908+0909 &  336 &  468 & 452 & 0.22 &  48 & 1.7 &N\\
J1910+0714 & 2712 &  124 & 106 & 0.36 &  14 & 1.8 &Y\\
B1913+10   &  404 &  242 & 240 & 1.3  &  30 & 4.4 &Y\\\\
J1913+1000 &  837 &  422 & 452 & 0.53 &  26 & 1.7 &Y\\
J1913+1011 &   35 &  179 & 170 & 0.50 &  10 & 2.7 &N\\
B1914+13   &  282 &  237 & 219 & 1.2  & 150 & 1.8 &Y\\
B1915+13   &  195 &   95 & 103 & 1.9  &  74 & 2.3 &Y\\
B1916+14   & 1181 &   27 & 28  & 1.0  &  18 & 3.0 &Y\\\\
B1919+14   &  618 &   92 &  74 & 0.7  &  41 & 0.5 &Y\\
B1921+17   &  547 &  143 & 177 & ---  &  13 & 3.0 &N\\
B1925+188  &  298 &   99 & 166 & ---  &  19 & 1.9 &N\\
B1929+20   &  268 &  211 & 205 & 1.2  &  19 & 3.0 &N\\
B1952+29   & 427  &    8 & 18  & 8.0  &  88 & 3.7 &Y\\\\
J1957+2831 & 308  &  139 & 163 & 1.0  &  54 & 1.6 &Y\\
B2000+32   & 697  &  142 & 184 & 1.2  &  43 & 2.2 &Y\\
J2002+30   & 422  &  196 & 184 & ---  &  24 & 1.2 &N\\
B2002+31   & 2111 &  235 & 197 & 1.8  &  88 & 3.3 &Y\\
\enddata
\tablecomments{Pulsar parameters $P$, DM and $S_{1400}$ are 
from the ATNF pulsar database \citep{mhth05}.
$\widehat{\rm DM}$ is the DM value at which the search algorithm identified
the pulsar.
$\langle S/N \rangle$ =
signal-to-noise ratio of the averaged pulse shape.
$\Delta\theta$ is the  angular distance from the nearest beam centroid in which
the pulsar was detected. SP  denotes whether 
individual pulses from this pulsar
were detected in the single-pulse search.
}
\end{deluxetable}

\begin{deluxetable}{ll}
\tablewidth{0pt}
\tablecaption{\label{tab:parms}Observed and derived parameters for \psrone}
\tablecolumns{2}
\tablehead{
\colhead{Parameter} &
\colhead{Value}
}
\startdata
Right ascension (J2000)\dotfill&$19^{\rm h}28^{\rm m}42\fs48(4)$\\
Declination (J2000)\dotfill    & $17\arcdeg46\arcmin27(1)\arcsec$ \\
Spin period, $P$ (ms)\dotfill                   & 68.728784754(1)        \\
Period derivative, $\dot P$\dotfill             & $1.3209(5)\times 10^{-14}$  \\
Epoch (MJD)\dotfill                                 & 53448.0               \\
Dispersion measure, DM (pc cm$^{-3}$)\dotfill      & 176.9(4)              \\
Flux density at 1400\,MHz, $S_{1400}$ (mJy)\dotfill & 0.25                  \\
\ \\[1pt] 
\tableline
\ \\[1pt] 
Surface magnetic field, $B$ (Gauss)\dotfill         & $9.6\times 10^{11}$ \\
Characteristic age, $\tau_c$ (kyr)\dotfill              & 82                \\
Spin-down luminosity, $\dot E$ (ergs\,s$^{-1}$)\dotfill& $1.6\times10^{36}$  \\
DM Distance (NE2001), $D$ (kpc)  \dotfill            & $\sim 6$          \\
Radio luminosity at 1400\,MHz, $S_{1400}D^2$ (mJy\,kpc$^2$)\dotfill &$\sim9$\\
\enddata
\tablecomments{
\noindent
Since the timing data collected so far span only 257 days,
the phase-connected timing solution should be viewed 
as preliminary. 
The figures in parentheses give the
uncertainties in the least-significant digits quoted.  To be conservative,
these  are calculated by
multiplying the nominal 1\,$\sigma$ {\sc tempo} standard 
deviations by an ad-hoc factor of 10. The DM distance is calculated
using the NE2001 electron density model for the Galaxy \citep{cl02}. 
}
\end{deluxetable}

\end{document}